\documentclass[conference,a4paper]{IEEEtran}

\usepackage[utf8]{inputenc} 
\usepackage[T1]{fontenc}
\usepackage{url}
\usepackage{ifthen}
\usepackage{cite}
\usepackage[cmex10]{amsmath} 
\usepackage{amssymb}
\usepackage[linesnumbered,ruled]{algorithm2e}
\usepackage{graphicx}
\usepackage{multirow}
\usepackage{makecell}

\newtheorem{exmp}{Example}
\newtheorem{cons}{Construction}

\begin{document}

\title{On Pliable Index Coding}
\author{\IEEEauthorblockN{Shanuja Sasi and B. Sundar Rajan} 
	\IEEEauthorblockA{Dept. of Electrical Communication Engg., Indian Institute of Science Bangalore, Karnataka, India - 560012\\
		Email: \{shanuja,bsrajan\}@iisc.ac.in}}	
\maketitle
\thispagestyle{plain}
\pagestyle{plain}

\begin{abstract}
	A new variant of index coding problem termed as Pliable Index Coding Problem (PICOD) is formulated in [S. Brahma, C. Fragouli, "Pliable index coding", IEEE Transactions on Information Theory, vol. 61, no. 11, pp. 6192-6203, 2015]. In PICOD, we consider a server holding a set of messages and there is a set of clients having a subset of messages with them. Each client is satisfied if it receives any of the message which it doesn't have. We discuss about a class of PICOD where the side information is consecutive. We provide index codes for two extreme cases -  for the class where each client gets exactly one desired message and  for a class where total number of messages decoded by the effective clients is maximized. Another variant of index coding problem is - c-Constrained Pliable Index Coding Problem [Linqi Song, Christina Fragouli and Tianchu Zhao, "A Pliable Index Coding Approach to Data Shuffling," arXiv:1701.05540v3 [cs.IT] 3 May 2018]. It is basically PICOD with a c-constraint, i.e, each message is decoded by atmost c clients demanding that message. We provide index codes for some classes of this variant with consecutive side information.
\end{abstract}

\section{Preliminaries}

In classical Index Coding Problem (ICP) \cite{bk,Birk}, there is a single server having a set of $P$ messages and is connected to a broadcast channel to a set of $N$ receivers or clients. Each client knows some subset of $P$ messages apriori which is termed as the side information and demands some other subset of messages which it doesn't have. The aim is to minimize the number of broadcast transmissions by the server such that it satisfies the requirements of each client. The index coding problem is a distributed source coding problem with side information that
has received considerable attention over the past decade. It was motivated by applications such as audio and video-on-demand, in satellite communication, in coded caching etc.

A new variant of the index coding problem was formulated in \cite{SBCF}, which is termed as Pliable Index Coding Problem (PICOD). In PICOD, the setting is same as that of ICP, i.e., there will be a server holding a set of $P$ messages and a set of $N$ clients or receivers. Each client knows some subset of $P$ messages apriori. What makes PICOD different from ICP is that in ICP each client demands a specific message from the server, while in PICOD each client is satisfied if it receives any message it does not have. The objective is to find the minimum number of coded transmissions that the server can make over a noiseless channel so that requirements of all the clients are met. It is motivated from its applications such as in Internet searching. For example, we are searching for latest news and we already have some information with us. We are happy if we get any additional news that we do not have, with minimum delay. Here, we are not specifying the news. This is exactly what happens in PICOD. PICOD($g$) represents a pliable index coding problem where the clients are satisfied it they receive any $g$ messages that they do not have.

 In \cite{SBCF}, an upper bound on the number of broadcast messages for solving any instance of PICOD(g) is obtained. In general, finding the optimal linear code for PICOD is NP-Hard, which is proved in . It is proved that there is an exponential improvement over the worst case scenario in index coding when the cardinalities of side information set are equal. A polynomial time heuristic approximation algorithms for solving PICOD($g$) is also provided.   

 In this paper we deal with two extreme cases of PICOD. One is where each client gets exactly one message, i.e, no clients get more than one message they have demanded. This has got application in media service providers where we actually pay for the movies. Service providers have a set of movies with them and the clients demands some movies. Clients pay for only certain number of movies, say one movie. Clients are satisfied if they get any movie they do not have. There is a restriction from the service providers' side as the clients have paid for only one movie. So they have a restriction that they can provide only one movie to each client. The clients are satisfied if they get any movie they have demanded. 
 
 The other extreme case is where with minimum delay the clients get maximum number of messages. Consider the above mentioned example where we are searching for latest news and we already have some information with us. In general case for PICOD we are happy if we get any additional news that we do not have with minimum delay. Here we are taking the case where we get maximum amount of information with the same delay.

 Let $\mathcal{M} = \{x_{0},x_{1},...,x_{P-1}\}$ be the set of all messages that the server holds and $\mathcal{R} =\{\mathcal{R}_{0},\mathcal{R}_{1},...,\mathcal{R}_{N-1} \}$ be the set of all clients or receivers. Throughout this paper, the message indices are taken modulo $P$. In this paper we consider a class of PICOD where each client has any $k$ consecutive messages as side information. The set of all possible side information patterns is $\mathcal{K} = \{\mathcal{K}_0, \mathcal{K}_1,...,\mathcal{K}_{P-1}\}$, where $\mathcal{K}_i =\{x_{i-1}, x_{i-2}, x_{i-3},...,x_{i-k}\}$, for each $i \in [0,P-1] $. Let $\mathcal{W}_i = \mathcal{M} \backslash \mathcal{K}_i$. 

Without loss of generality, throughout this paper the clients with same set of side information are treated alike. So effectively there are only $P$ clients since the total number of side information patterns possible is $P$. Throughout this paper when we refer to clients, it is basically effective clients. 

Let $\mathcal{C}_i$ represent the clients whose side information set is $\mathcal{K}_i$, where $i \in [0,P-1]$. Let $\mathcal{C}=\{\mathcal{C}_0, \mathcal{C}_1,...,\mathcal{C}_{P-1}\}$, which represent the effective clients. 

In this paper we take two extreme cases.

(1) Each client gets exactly one message.

%(2) Let $n_i$ be the number of messsages client $\mathcal{C}_{i}$ get from the coded transmissions, where $i \in [0,P-1]$. We take the case where $\sum_{i=0}^{P-1} n_i$ is maximum.

(2) Total number of messages decoded by the effective clients is maximized.

Another variant of index coding is Constrained Pliable Index Coding. In c-Constrained Pliable Index Coding, each message is decoded by atmost $c$ clients who demand that message. We put a restriction on the number of clients who can a decode a message. In this paper, we discuss about a class of PICOD where each client has any $k$ consecutive messages as side information, under a c-constraint, i.e., each message is decoded by atmost c clients demanding that message.

\textit{Notations:} $[a,b]$ represent the set of integers $\{a,a+1,a+2,...,b\}$. $ \lceil a \rceil$ represent the integer greater than or equal to $a$. $ \lfloor a \rfloor$ represent the integer less than or equal to $a$. 

\subsection{Our Contributions}
The contributions in this paper is summarized as follows.
\begin{itemize}
	\item We provide index code for a class of PICOD where side information is consecutive and each client gets exactly one desired message in Section \ref{sec:case1}.
	\item In Section \ref{sec:case2}, we give index code for a class of PICOD where total number of messages decoded by the effective clients is maximized.
	
	\item In Section \ref{sec:case2}, we prove that the index code provided for a class of PICOD, where total number of messages decoded by the effective clients is maximized, is optimal. 
	\item We discuss about c-constrained pliable index coding with consecutive side information. We provide index code for such cases in Section \ref{Sec:Case3}. 
\end{itemize}
 
\section{Each client gets exactly one message.}
\label{sec:case1}
In this section we will provide index code for the first extreme case where each client gets exactly one message. We will prove that for some values of $k$, there doesn't exist a case where each client gets exactly one message, i.e, atleast one client gets more than one message whatever be the coded transmission.

\textit{Case 1:} For $3 \leq k < \lceil \frac{P}{2} \rceil$.

With $\lceil \frac{P-k-1}{k-1} \rceil$ number of transmissions, each client gets exactly one message if $3 \leq k < \lceil \frac{P}{2} \rceil$. The code construction is as follows.

\begin{cons}

Let $T = \lceil \frac{P-k-1}{k-1} \rceil$. Let $y=P-2k-1$ and $r = y$ mod $ (k-1)$. 

\begin{itemize}
	\item Pick some integer $i \in [0,P-1]$.
	\item A coded symbol is obtained by XOR of the messages $x_i$ and $x_{i+P-k}$, i.e, $$w_1 = x_i \oplus x_{i+P-k}$$.
	\item If $T \geq 3$, do the following. Let $V'_2 = k-1$  and $V''_2 = V'_2 +1$. For each $ l \in [3,T-1]$, let 
	\begin{align*}
	V'_l &= V''_{l-1} +k-2 \\ V''_l &= V''_{l-1} +k-1 .
	\end{align*}  For each $ j \in [2,T-1]$, a coded symbol $w_j$ is obtained, where
	
	\begin{equation*}
	w_j=
	\begin{cases}
	x_{i+V'_j} \oplus x_{i+V''_j} \oplus x_{i+j-k-1} , & \text{if}\ j \leq k \\
	x_{i+V'_j} \oplus x_{i+V''_j} \oplus x_{i+j-k}, & \text{otherwise}.
	\end{cases}
	\end{equation*}
	
	\item Let $V_T^{(1)} = k$ if $T =2$, $V_T^{(1)} = V''_{T-1} +k-1$ if $T \geq 3$. For each $s \in [2,k-r]$, let $$V^{(s)}_T = V^{(s-1)}_T -1$$.
	A coded symbol $w_T$ is obtained, where
	\begin{equation*}
	w_T=
	\begin{cases}
	x_{i+T-k-1} \oplus_{n=1}^{k-r} x_{i+V^{(n)}_T}, & \text{if}\ T \leq k. \\
	x_{i+T-k} \oplus_{n=1}^{k-r} x_{i+V^{(n)}_T}, & \text{otherwise}.
	\end{cases}
	\end{equation*}
	
\end{itemize}

\end{cons}

Let

\begin{equation*}
\chi_j =
\begin{cases}
\{\mathcal{C}_{i-k+1}, \mathcal{C}_{i-k+2},..,\mathcal{C}_{i+k}\}, & \text{for}\ j =1  \\
\{\mathcal{C}_{i+(j-1)k+3-j},..,\mathcal{C}_{i+jk-j+1}\}, & \text{for}\ j \in [2,T-1] \\
\{\mathcal{C}_{i+(T-1)k+3-T},..,&\\ \mathcal{C}_{i+(T-1)k+3-T+r}\}, & \text{for}\ j =T

\end{cases}
\end{equation*}

\begin{table*}
	\begin{center}
		\begin{tabular}{ | p{3cm} | p{2.9cm} | p{3cm}  |}
			\hline
			
			\textit{Clients}& \textit{The message $x_l$ decoded by the clients}  &  \textit{The coded symbol from which the message $x_l$ is decoded} \\ \hline
			$\{\mathcal{C}_{i-k+1}, \mathcal{C}_{i-k+2},..,\mathcal{C}_{i}\}$ & $x_i$ & $w_1$ \\ \hline
			$\{\mathcal{C}_{i+1}, \mathcal{C}_{i+2},..,\mathcal{C}_{i+k}\}$ & $x_{i+P-k}$ & $w_1$ \\ \hline
			$\chi_j$, for each $j \in [2,T]$ and $j \leq k$ & 
			$x_{i+j-k-1}$ & $w_j$ \\ \hline
			$\chi_j$, for each $j \in [2,T]$ and $j > k$ & 
			$x_{i+j-k}$ & $w_j$ \\ \hline
			
		\end{tabular}
	\end{center}
	\caption{Table that illustrates the messages decoded by the clients from the coded symbols transmitted for $q=1$.}
	\label{tab1}
\end{table*}

\begin{table*}
	\begin{center}
		\begin{tabular}{ | p{2.5cm} | p{3cm} | p{4.2cm}  |p{4.5cm}|}
			\hline
			
			\textit{Coded symbol $w_l$}& \textit{Client $C_l$ who decoded some message $x_l$ from $w_l$}  &  \textit{Clients other than $C_l$} & \textit{XOR of a subset of messages $\mathcal{W}_i \backslash x_l$ present in $w_l$} \\ \hline
			$w_1$ & $\chi_{1}$ & $\mathcal{C} \backslash \chi_{1}$ & $x_i \oplus x_{i+P-k}$ \\ \hline
			$w_j$, for each $j \in [2,T-1]$ & $\chi_{j}$ &
			
			\multirow{3}{*}{$\{\mathcal{C}_{i-k+j},...,\mathcal{C}_{i+j-1}\}$} &$x_{i+V'_j} \oplus x_{i+V''_j}$ \\ \cline{3-4}
			& &$\{\mathcal{C}_{i+j},...,\mathcal{C}_{i+jk-k-j+2}\}$ & \makecell{$x_{i+j-k-1} \oplus x_{i+V''_j}$, if $j \leq k.$ \\$x_{i+j-k} \oplus x_{i+V''_j}$,    otherwise.}\\ \cline{3-4}
			&& $\{\mathcal{C}_{i+jk-j+2},...,\mathcal{C}_{i+j-k-1}\}$ & \makecell{$	x_{i+j-k-1},x_{i+V'_j}$, if $j \leq k$ \\
				$x_{i+j-k},x_{i+V'_j},$ otherwise.} \\ \hline
			
			$w_T$ & $\chi_{T}$ &
			
			\multirow{4}{*}{$\{\mathcal{C}_{i-k+T},...,\mathcal{C}_{i+T-1}\}$ } &$x_{i+V^{(1)}_T} \oplus x_{i+V^{(2)}_T}$ \\ &&&\\ \cline{3-4}
			& &$\{\mathcal{C}_{i+T},...,\mathcal{C}_{i+Tk-k-T+2}\}$ & \makecell{$x_{i+T-k-1} \oplus x_{i+V^{(1)}_T},$ if $T \leq k$. \\
				$x_{i+T-k} \oplus x_{i+V^{(1)}_T}$, otherwise}\\ \cline{3-4}
			&& $\{\mathcal{C}_{i+Tk-k-T+4+r},...,\mathcal{C}_{i+T-k-1}\}$ & \makecell{$x_{i+T-k-1} \oplus x_{i+V^{(k-r)}_T}$, if $T \leq k$ \\
				$x_{i+T-k} \oplus x_{i+V^{(k-r)}_T}$, otherwise} \\ \hline

		\end{tabular}
	\end{center}
	\caption{Table that illustrates the uniqueness of the messages decoded by the clients discussed in Case 1.}
	\label{tab2}
\end{table*}

Table \ref{tab1} illustrates the message decoded by each client and the coded symbol from which it has decoded that message. If XOR of more than one unknown message (to a client) is present in a coded symbol, then that client cannot decode any message from that coded symbols. For every client, in all other coded symbols except the one from which it decoded one message (as illustrated in Table \ref{tab1}), XOR of more than one unknown message (to a client) is present (It is illustrated in table \ref{tab2}).  Hence each client cannot retrieve more than one message.

\begin{exmp}
	Let $P=7$ and $k=3$. Here $T = \lceil \frac{P-k-1}{k-1} \rceil =2.$ Hence two transmissions are required. Here $y=P-2k-2 =0$ and $r = 0$. 
	Following the procedures as in \textit{Construction 1}:
	
	\begin{itemize}
		\item Let $i=3$.
		\item A coded symbol $w_1$ is obtained where $w_1 = x_3 \oplus x_0$.
		\item $k-r=3$. Hence $V_T^{(1)}=3, V_T^{(2)}=2,V_T^{(3)}=1$, since $T=2$. A coded symbol $w_2$ is obtained where $w_2= x_4 \oplus x_5 \oplus x_6$.
	\end{itemize}

	 $\{\mathcal{C}_4,\mathcal{C}_5,\mathcal{C}_6\}$ get $x_0$ and the clients in $\{\mathcal{C}_1,\mathcal{C}_2,\mathcal{C}_3\}$ get $x_3$ from $w_1$. The clients in $\{\mathcal{C}_0\}$ get $x_1$ from $w_2$. Since $x_4 \oplus x_5$ is present in $w_2$, the clients in $\{\mathcal{C}_2,\mathcal{C}_3,\mathcal{C}_4\}$ cannot retrieve any message from $w_2$ as $x_4$ and $x_5$ are not present as side information with those clients. Similarly the clients in $\{\mathcal{C}_5,\mathcal{C}_6\}$ cannot retrieve any message from $w_2$ since $x_1 \oplus x_6$ is present in $w_2$. The clients in $\mathcal{C}_1$ cannot get any message from $w_2$ as $x_1 \oplus x_4$ is present in $w_2$ while the clients in $\mathcal{C}_0$ cannot get any message from $w_1$ as $x_3 \oplus x_0$ is present in $w_1$. Hence all the clients get exactly one message.
\end{exmp}

\textit{Case 2:} $\lfloor \frac{P}{2}  \rfloor< k \leq P-4$.

Let us take the case where $\lfloor \frac{P}{2}  \rfloor< k \leq P-4$. If $P$ is divisible by $P-k$, with one coded transmission each client gets one message it doesn't have. For all other values of $k$, two transmissions are required. The code construction is given below for such cases.

\begin{cons}
Let $P=t(P-k)+q$. 

\begin{itemize}
	\item Pick some integer $i \in [0,P-1]$.
	\item If $q=0$, a coded symbol $w_1$ is obtained, where  $$w_1 = \bigoplus_{g=0}^{t-1}x_{i+g(P-k)}$$.
	\item If $q=1$, 
	\begin{itemize}
		\item Two coded symbols $w_1$ and $w_2$ are obtained, where 
		\begin{align*}
		w_1&=\bigoplus_{g=0}^{t}x_{i+g(P-k)}.\\
		w_2 &=
		\begin{cases}
		x_{i+1} \oplus x_{i+P-k-1}\\ \oplus \left(\bigoplus_{b=1}^{2} x_{i+(t-1)(P-k)+b}\right), & \text{if}\ t=2. \\
		x_{i+1} \oplus x_{i+P-k-1}\\ \oplus \left(\bigoplus_{b=1}^{2} x_{i+(t-1)(P-k)+b}\right) \oplus \\ \left(\bigoplus_{\substack{b \in [1,t-2],\\a \in [1,P-k-1]}} x_{i+b(P-k)+a}\right), & \text{if}\ t \geq 3.
		\end{cases}
		\end{align*}
	\end{itemize}
	
	\item If $q \geq 2$,
	\begin{itemize}
		\item Two coded symbols $w_1$ and $w_2$ are obtained, where 
		\begin{align*}
				w_1&=\bigoplus_{g=0}^{t}x_{i+g(P-k)}.\\
				w_2 &=
				\begin{cases}
				x_{i-q+1} \oplus \left(\bigoplus_{b=1}^{q} x_{i+(P-k)-b}\right) \\
				\oplus \left(\bigoplus_{b=1}^{q} x_{i+(t-1)(P-k)+b}\right), 
				& \text{if}\ t=2. \\
				x_{i-q+1} \oplus \left(\bigoplus_{b=1}^{q} x_{i+(P-k)-b}\right) \\
				 \oplus \left(\bigoplus_{b=1}^{q} x_{i+(t-1)(P-k)+b}\right) \oplus\\
				 \left(\bigoplus_{ \substack{b \in [1,t-2]\\a \in [1,P-k-1]}} x_{i+b(P-k)+a}\right), 
				& \text{if}\ t \geq 3.
				\end{cases}
	\end{align*}

	\end{itemize}

\end{itemize}
\end{cons}

\begin{table*}
	\begin{center}
		\begin{tabular}{ |p{1.2cm}| p{4.5cm} | p{2.9cm} | p{3.5cm}  |}
			\hline
			
		\textit{Value of q}	&\textit{Clients}& \textit{The message $x_l$ decoded by the clients}  &  \textit{The coded symbol from which the message $x_l$ is decoded} \\ \hline
		$0$&	$\{\mathcal{C}_j\}$, where $j \in [i+(g)(P-k)+1,i+(g+1)(P-k)],g \in [0,t-1]$ & $x_{i+(g+1)(P-k)}$ & $w_1$ \\ \hline
		$1$&	$\mathcal{C}_i$ & $x_i$ & $w_1$ \\ \cline{2-4}
			&$\{\mathcal{C}_{i+g(P-k)+1},...,\mathcal{C}_{i+(g+1)(P-k)}\}$, for each $g \in [0,T-2]$ & $x_{i+(g+1)(P-k)}$ & $w_1$ \\ \cline{2-4}
			&$\mathcal{C}_{i+(t-1)(P-k)+1}$ & $x_{i+t(P-k)}$ & $w_1$ \\ \cline{2-4}
			&
			$\mathcal{C}_{i+(t-1)(P-k)+2}$ & $x_{i+(t-1)(P-k)+2}$  & $w_2$ \\ \cline{2-4}
			
			&$\{\mathcal{C}_{i+(t-1)(P-k)+3},...,\mathcal{C}_{i+t(P-k)}\}$ & $x_{i+1}$ & $w_2$ \\ \hline
			
		\end{tabular}
	\end{center}
	\caption{Table that illustrates the messages decoded by the clients from the coded symbols transmitted for $q=0,1$ (Case 2).}
	\label{tab3}
\end{table*}

\begin{table*}
	\begin{center}
		\begin{tabular}{ | p{2cm} | p{4.2cm} | p{4.2cm}  |p{4.5cm}|}
			\hline
			
			\textit{Coded symbol $w_l$}& \textit{Client $C_l$ who decoded some message $x_l$ from $w_l$}  &  \textit{Clients other than $C_l$} & \textit{XOR of a subset of messages $\mathcal{W}_i \backslash x_l$ present in $w_l$} \\ \hline
			
			$w_1$ & $\{\mathcal{C}_i,...,\mathcal{C}_{i+(t-1)(P-k)+1}\}$ &$\{\mathcal{C}_{i+(t-1)(P-k)+2},...,\mathcal{C}_{i+t(P-k)}\}$ & $x_{i} \oplus x_{i+t(P-k)}$\\ \hline
			$w_2$ & $\{\mathcal{C}_{i+(t-1)(P-k)+2},...,\mathcal{C}_{i+t(P-k)}\}$ & $\{\mathcal{C}_{i+2},...,\mathcal{C}_{i+P-k-1}\}$& $x_{i+P-k-1} \oplus x_{i+P-k+1}$ \\  \cline{3-4}
			
			\multirow{6}{*}{}
			& &$\{\mathcal{C}_{i},\mathcal{C}_{i+1}\}$ & $x_{i+1} \oplus x_{i+P-k-1}$\\ \cline{3-4}
			&& $\{\mathcal{C}_j\}$, where $j \in (i+g(P-k)),g \in [1,t-1]$ & $x_{j+1} \oplus x_{j+2}$ \\ \cline{3-4}
			&& $\{\mathcal{C}_j\}$, where $j \in \{i+g(P-k)+1,...,i+(g+1)(P-k)-2\},g \in [1,t-2]$ (if $t \geq 3$) & $x_{j} \oplus x_{j+1}$ \\ \cline{3-4}
			&& $\{\mathcal{C}_j\}$, where $j \in (i+(g+1)(P-k)-1),g \in [1,t-2]$ (if $t \geq 3$) & $x_{j} \oplus x_{j+2}$ \\ \cline{3-4}
			
			&& $\{\mathcal{C}_j\}$, where $j= i+(t-1)(P-k)+1$  & $x_{j} \oplus x_{j+1}$	\\ \hline

		\end{tabular}
	\end{center}
	\caption{Table that illustrates the uniqueness of the messages decoded by the clients for $q=1$ (Case 2).}
	\label{tab4}
\end{table*}

 \begin{table*}
	\begin{center}
		\begin{tabular}{ | p{5cm} | p{2.9cm} | p{3.5cm} |}
			\hline
			
			\textit{Clients}& \textit{The message $x_l$ decoded by the clients}  &  \textit{The coded symbol from which the message $x_l$ is decoded} \\ \hline
			$\{\mathcal{C}_{i+g(P-k)+1},...,\mathcal{C}_{i+(g+1)(P-k)}\}$, for $g \in [0,t-2]$  & $x_{i+(g+1)(P-k)}$ &$w_1$\\ \hline
				$\{\mathcal{C}_{i-q+1},...,\mathcal{C}_{i}\}$  & $x_{i}$ &$w_1$\\ \hline
				
					$\{\mathcal{C}_{i+(t-1)(P-k)+1},...,\mathcal{C}_{i+(t-1)(P-k)+q}\}$& $x_{i+(t)(P-k)}$ &$w_1$\\ \hline
				
			$\{\mathcal{C}_{i+(t-1)(P-k)+q+1},...,\mathcal{C}_{i+t(P-k)}\}$ & $x_{i-q+1}$ & $w_2$ \\ \hline
			
		\end{tabular}
	\end{center}
	\caption{Table that illustrates the messages decoded by the clients from the coded symbols transmitted for $q \geq 2$ (Case 2).}
	\label{tab5}
\end{table*}

\begin{table*}
	\begin{center}
		\begin{tabular}{ | p{1.8cm} | p{4.5cm} | p{4.8cm}  |p{4cm}|}
			\hline
			
			\textit{Coded symbol $w_l$}& \textit{Client $C_l$ who decoded some message $x_l$ from $w_l$}  &  \textit{Clients other than $C_l$} & \textit{XOR of a subset of messages $\mathcal{W}_i \backslash x_l$ present in $w_l$} \\ \hline
			
			$w_1$ & $\{\mathcal{C}_{i+t(P-k)+1},...,\mathcal{C}_{i+(t-1)(P-k)+q}\}$ &$\{\mathcal{C}_{i+(t-1)(P-k)+q+1},...,\mathcal{C}_{i+t(P-k)}\}$ & $x_{i} \oplus x_{i+t(N-k)}$\\ \hline
			& & $\{\mathcal{C}_{i-q+1}\}$ & $x_{i-q+1} \oplus x_{i+P-k-q}$ \\  \cline{3-4}
			
			\multirow{6}{*}{}&& $\{\mathcal{C}_{i-q+2},...,\mathcal{C}_{i+P-k-q}\}$ & $x_{i+P-k-q} \oplus x_{i+P-k-q+1}$ \\ \cline{3-4}
			& &$\{\mathcal{C}_j\}$, where $j \in [i+P-k-q+1,i+P-k-2]$ & $x_{j} \oplus x_{j+1}$\\ \cline{3-4}
			$w_2$ & $\{\mathcal{C}_{i+(t-1)(P-k)+q+1},...,\mathcal{C}_{i+t(P-k)}\}$& $\{\mathcal{C}_j\}$, where $j \in [i+g(P-k)+1,i+(g+1)(P-k)-2],g \in [1,t-2]$ (if $t \geq 3$) & $x_{j} \oplus x_{j+1}$ \\ \cline{3-4}
			&& $\{\mathcal{C}_j\}$, where $j \in \{i+(g+1)(P-k)-1\},g \in [0,t-2]$ & $x_{j} \oplus x_{j+2}$ \\ \cline{3-4}
			
			&& $\{\mathcal{C}_j\}$, where $j \in \{i+g(P-k)\},g \in [1,t-1]$ & $x_{j+1} \oplus x_{j+2}$ \\ \cline{3-4}
			
			&& $\{\mathcal{C}_j\}$, where $j \in [i+(t-1)(P-k)+1,i+(t-1)(P-k)+q-1]$ & $x_{j} \oplus x_{j+1}$ \\ \cline{3-4}
			
			&& $\{\mathcal{C}_{i+(t-1)(P-k)+q}\}$ & $x_{i-q+1} \oplus x_{i+(t-1)(P-k)+q}$	\\ \hline

		\end{tabular}
	\end{center}
	\caption{Table that illustrates the uniqueness of the messages decoded by the clients for $q \geq 2$ (Case 2).}
	\label{tab6}
\end{table*}

Table \ref{tab3} and \ref{tab5} illustrate the message decoded by each client and the coded symbol from which it is decoded for $q=0,1$ and $q \geq 2 $ respectively. It is illustrated in table \ref{tab4} and \ref{tab6} that each client doesn't get more than one message for $q=1$ and $q \geq 2 $ respectively (for the same reason as in Case 1).

%\begin{table*}
%	\begin{center}
%		\begin{tabular}{| p{4.2cm}  |p{4.5cm}|}
%			\hline
%			
%			\textit{Coded symbol $w_l$}& \textit{Client $C_l$ who decoded some message $x_l$ from $w_l$}  \\ \hline
%			
%			$k=1$ and $P$ even & $w_j = x_{i+2j-2} \oplus x_{i+2j-1}$, for each $j \in [1,\frac{P}{2}]$.\\ \hline
%			
%			$k=P-1$ & $w_1= \bigoplus_{i=0}^{P-1} x_i$ \\ \hline
%			
%			$k=P-2$ and $P$ even & $w_1=\bigoplus_{i=0}^{\frac{P-2}{2}} x_{2i}\}$ \\ \hline
%			
%			$k=P-3$ and $P$ odd & $w_1 = \bigoplus_{i=0}^{\frac{P-1}{2}} x_{2i}$ \\ 
%			& $w_2 = \bigoplus_{i=0}^{\frac{P-1}{2}-1} x_{2i+1}$\\ \hline
%			
%			$k=P-3$ and $P$ even & $w_1 = \bigoplus_{i=0}^{\frac{P-2}{2}} x_{2i}$ \\ 
%			& $w_2 = \bigoplus_{i=0}^{\frac{P-2}{2}} x_{2i+1}$\\ \hline			
%			
%		
%			
%		\end{tabular}
%	\end{center}
%	\caption{Table that illustrates the index code for $k = 1,N-1,N-2,N-3$.}
%	\label{tab8}
%\end{table*}

\begin{exmp}
	Let $P=9$ and $k=5$. Here $t=2$, $q=1$. 
	Following the procedure as in \textit{Construction 1},
	\begin{itemize}
		\item Let $i=1$.
		\item A coded symbol $w_1$ is obtained where $w_1 = x_1 \oplus x_5 \oplus x_0$.
		\item A coded symbol $w_2$ is obtained where $w_2= x_2 \oplus x_4 \oplus x_6 \oplus x_7$.
	\end{itemize}
 The clients in $\{\mathcal{C}_2,\mathcal{C}_3,\mathcal{C}_4, \mathcal{C}_5\}$ get $x_7$, those in $\{\mathcal{C}_1\}$ get $x_1$ and those in $\{\mathcal{C}_6\}$ get $x_0$ from $w_1$. The clients in $\{\mathcal{C}_7\}$ get $x_7$ while the clients in $\{\mathcal{C}_8, \mathcal{C}_0\}$ get $x_2$ from $w_2$. Since $x_2 \oplus x_4$ is present in $w_2$, the clients in $\{\mathcal{C}_1,\mathcal{C}_2\}$ cannot retrieve any message from $w_2$. Similarly the clients in $\{\mathcal{C}_3,\mathcal{C}_4\}$ cannot retrieve any message from $w_2$ since $x_4 \oplus x_6$ is present in $w_2$. The clients in $\{\mathcal{C}_5,\mathcal{C}_6\}$ cannot get any message from $w_2$ as $x_6 \oplus x_7$ is present in $w_2$ while the clients in $\{\mathcal{C}_7,\mathcal{C}_8, \mathcal{C}_0\}$ cannot get any message from $w_1$ as $x_1 \oplus x_0$ is present in $w_1$. Hence all the clients get exactly one message.
\end{exmp}

\textit{Case 3:} $k=1,2,P-1,P-2,P-3$.

 For $k=1$ and $P$ even, the total number of transmission required is $\frac{P}{2}$. For $k=P-2$ and $P$ even, with just one coded transmission each client gets one message not in the side information set while for $k=P-3$ and $P$ even, two transmissions are required. It is trivial for $k=P-1$, where the coded transmission is the sum of all messages the server has. When it comes to $k=2,$ for different values of $q$, the number of transmissions required also varies.  The code construction for all such cases is given below.

\begin{cons}
	Let $P=t(P-k)+q$ if $k \geq \lfloor \frac{P}{2}  \rfloor$. If $k=2,$ then let $P=4t+q$. 

\begin{itemize}
	\item Pick some integer $i \in [0,P-1]$.
	\item For $k=P-1$, a coded symbol $w_1$ is obtained, where $$w_1= \bigoplus_{i=0}^{P-1} x_i$$.
	\item For $k=1$ and $P$ even, a coded symbol $w_j$ is obtained for each $j \in [1,\frac{P}{2}]$, where $$w_j = x_{i+2j-2} \oplus x_{i+2j-1}$$.
	\item For $k=2$, 
	\begin{itemize}
		\item If $q=0$, $t$ coded symbols are obtained, i.e, for each $g=[0,t-1]$,
		\begin{align*}
		w_{g+1}=x_{i+4g} \oplus x_{i+4g+2}.
		\end{align*}
		\item If $q=1,t+1$ coded symbols are obtained, i.e,
		\begin{align*}
		w_{g+1}&=x_{i+4g} \oplus x_{i+4g+2}, \text{for each $g=[0,t-1]$,} \\
		w_{t+1} &= x_{i+1} \oplus x_{i+4t-1} \oplus x_{i+4t}.
		\end{align*}
		\item If $q=2,t+1$ coded symbols are obtained, i.e,
		\begin{align*}
		w_{g+1}&=x_{i+4g} \oplus x_{i+4g+2}, \text{for each $g=[0,t-1]$,} \\
		w_{t+1} &= x_{i+4t-1} \oplus x_{i+4t} \oplus x_{i+4t+1}.
		\end{align*}
		\item If $q=2,t+2$ coded symbols are obtained, i.e,
		\begin{align*}
		w_{g+1}&=x_{i+4g} \oplus x_{i+4g+2}, \text{for each $g=[0,t-1]$,} \\
		w_{t+1} &= x_{i+4t-1} \oplus x_{i+4t} \oplus x_{i+4t+1},\\
		w_{t+2} &= x_{i+1} \oplus x_{i+4t+1} \oplus x_{i+4t+2}.
		\end{align*}
			
	\end{itemize}

	\item For $k=P-2$ and $P$ even, a coded symbol $w_1$ is obtained, where $$w_1=\bigoplus_{i=0}^{\frac{P-2}{2}} x_{2i}$$.
	\item For $k=P-3$ and $P$ odd, if $q \neq 0$, two coded symbols are obtained, i.e, 
	\begin{align*}
	w_1 &= \bigoplus_{i=0}^{\frac{P-1}{2}} x_{2i},\\
	w_2 &= \bigoplus_{i=0}^{\frac{P-3}{2}} x_{2i+1}.
	\end{align*} 
	Else if $q=0$, a coded symbol $w_1$ is obtained, where  $$w_1 = \bigoplus_{g=0}^{t-1}x_{i+g(P-k)}$$.
	\item For $k=P-3$ and $P$ even, if $q \neq 0$, two coded symbols are obtained, i.e,
	\begin{align*}
	w_1 &= \bigoplus_{i=0}^{\frac{P-2}{2}} x_{2i}, \\
	w_2 &= \bigoplus_{i=0}^{\frac{P-2}{2}} x_{2i+1}.
	\end{align*}

	Else if $q=0$, a coded symbol $w_1$ is obtained, where  $$w_1 = \bigoplus_{g=0}^{t-1}x_{i+g(P-k)}$$.
\end{itemize}

\end{cons}

 \begin{table*}
	\begin{center}
		\begin{tabular}{ |p{1.5cm}| p{4.5cm} | p{2.9cm} | p{3.5cm} |}
			\hline
			
			 &\textit{Clients}& \textit{The message $x_l$ decoded by the clients}  &  \textit{The coded symbol from which the message $x_l$ is decoded} \\ \hline
			$k=1$ and $P$ even & $\{\mathcal{C}_{i+2j-1}\}$ for each $j \in [1,\frac{P}{2}]$ &  $x_{i+2j-1}$ & $w_j$\\ \hline

			& $\{\mathcal{C}_{i+2j}\}$ for each $j \in [1,\frac{P}{2}]$ &  $x_{i+2j-2}$ & $w_j$\\ \hline	
			
			$k=2$ and $q=0$& $\{\mathcal{C}_{i+4g+1},\mathcal{C}_{i+4g+2}\}$, for each $g \in [0,t-1]$ &$x_{i+4g+2}$& $w_{g+1}$ \\ \cline{2-4}
			&$\{\mathcal{C}_{i+4g+3},\mathcal{C}_{i+4g+4}\},$ for each $g \in [0,t-1]$&$x_{i+4g}$ &$w_{g+1}$ \\ \hline
			
			$k=2$ and $q=1$& $\{\mathcal{C}_{i+4g+1},\mathcal{C}_{i+4g+2}\}$, for each $g \in [0,t-1]$ &$x_{i+4g+2}$& $w_{g+1}$ \\ \cline{2-4}
			&$\{\mathcal{C}_{i+4g+3},\mathcal{C}_{i+4g+4}\},$ for each $g \in [0,t-1]$&$x_{i+4g}$ &$w_{g+1}$ \\ \cline{2-4}
			&$\{\mathcal{C}_{i}\}$ & $x_{i+1}$& $w_{t+1}$ \\ \hline
			
			$k=2$ and $q=2$& $\{\mathcal{C}_{i+4g+1},\mathcal{C}_{i+4g+2}\}$, for each $g \in [0,t-1]$ &$x_{i+4g+2}$& $w_{g+1}$ \\ \cline{2-4}
			&$\{\mathcal{C}_{i+4g+3},\mathcal{C}_{i+4g+4}\},$ for each $g \in [0,t-1]$&$x_{i+4g}$ &$w_{g+1}$ \\ \cline{2-4}
			&$\{\mathcal{C}_{i-1}\}$ & $x_{i+4t+1}$& $w_{t+1}$ \\ \cline{2-4}
			&$\{\mathcal{C}_{i}\}$ & $x_{i+4t-1}$& $w_{t+1}$ \\\hline
			
			$k=2$ and $q=3$& $\{\mathcal{C}_{i+4g+1},\mathcal{C}_{i+4g+2}\}$, for each $g \in [0,t-1]$ &$x_{i+4g+2}$& $w_{g+1}$ \\ \cline{2-4}
			&$\{\mathcal{C}_{i+4g+3},\mathcal{C}_{i+4g+4}\},$ for each $g \in [0,t-1]$&$x_{i+4g}$ &$w_{g+1}$ \\ \cline{2-4}
			&$\{\mathcal{C}_{i-2}\}$ & $x_{i+4t+1}$& $w_{t+1}$ \\  \cline{2-4}
			&$\{\mathcal{C}_{i-1}\}$ & $x_{i+4t-1}$& $w_{t+1}$ \\  \cline{2-4}
			&$\{\mathcal{C}_{i}\}$ & $x_{i+1}$& $w_{t+2}$ \\\hline
			
			$k=P-2$ and $P$ even & $\{\mathcal{C}_{j}\}$, where $j \in \{2g\}, g \in [0,\frac{P-2}{2}]$ & $x_j$ & $w_1$ \\ \cline{2-4}
			&  $\{\mathcal{C}_{j}\}$, where $j \in \{2g+1\}, g \in [0,\frac{P-2}{2}]$ & $x_{j+1}$ & $w_1$  \\ \hline
			
			 $k=P-3$ and $q=0$ & $\{\mathcal{C}_j\}$, where $j \in [i+(g)(P-k)+1,i+(g+1)(P-k)],g \in [0,t-1]$ & $x_{i+(g+1)(P-k)}$ & $w_1$ \\ \hline

		    $k=P-3$, $P$ odd and $q \neq 0$& $\{\mathcal{C}_{j}\}$, where $j \in \{2g\}, g \in [0,\frac{P-3}{2}]$ & $x_{j+1}$ & $w_2$ \\ \cline{2-4}
		    & $\{\mathcal{C}_{j}\}$, where $j \in \{2g+1\}, g \in [0,\frac{P-5}{2}]$ & $x_{j+1}$ & $w_1$ \\ 
		    & $\{\mathcal{C}_{j}\}$, where $j \in \{P-2\}$ & $x_{j}$ & $w_2$ \\ \cline{2-4}
		    & $\{\mathcal{C}_{j}\}$, where $j \in \{P-1\}$ & $x_{j+2}$ & $w_2$ \\ \cline{2-4}
		    \hline
		    $k=P-3$, $P$ even and $q \neq 0$ & $\{\mathcal{C}_{j}\}$, where $j \in \{2g\}, g \in [0,\frac{P-2}{2}]$ & $x_{j+1}$ & $w_2$ \\ \cline{2-4}
		    & $\{\mathcal{C}_{j}\}$, where $j \in \{2g+1\}, g \in [0,\frac{P-2}{2}]$ & $x_{j+1}$ & $w_1$ \\ \hline
			
		\end{tabular}
	\end{center}
	\caption{Table that illustrates the messages decoded by the clients from the coded symbols transmitted for $k=1,N-3$.}
	\label{tab10}
\end{table*}

\begin{table*}
	\begin{center}
		\begin{tabular}{ |p{1.5cm}| p{1.8cm} | p{4.5cm} | p{4.8cm}  |p{4cm}|}
			\hline
			
			&\textit{Coded symbol $w_l$}& \textit{Client $C_l$ who decoded some message $x_l$ from $w_l$}  &  \textit{Clients other than $C_l$} & \textit{XOR of a subset of messages $\mathcal{W}_i \backslash x_l$ present in $w_l$} \\ \hline
			
			$k=1$ and $P$ even &$w_j$, for $j \in [1,\frac{P}{2}]$ &$\{\mathcal{C}_{i+2j-1} \cup \mathcal{C}_{i+2j}\}$ &  $\mathcal{C} \backslash \{\mathcal{C}_{i+2j-1} \cup \mathcal{C}_{i+2j}\}$ & $x_{i+2j-2} \oplus x_{i+2j}$ \\ \hline
			
			$k=2$ and $q=0$& $w_{g+1},$ for $g \in [0,t-1]$ &$ \bigcup_{j=1}^{4} \{\mathcal{C}_{i+4g+j}\}$ &  $\mathcal{C} \backslash \bigcup_{j=1}^{4} \{\mathcal{C}_{i+4g+j}\}$ & $x_{i+4g} \oplus x_{i+4g+2}$ \\ \hline
			
			$k=2$ and $q=1$& $w_{g+1},$ for $g \in [0,t-1]$ &$ \bigcup_{j=1}^{4} \{\mathcal{C}_{i+4g+j}\}$ &  $\mathcal{C} \backslash \bigcup_{j=1}^{4} \{\mathcal{C}_{i+4g+j}\}$ & $x_{i+4g} \oplus x_{i+4g+2}$ \\ \cline{2-5}
			& $w_{t+1}$ & $\{\mathcal{C}_{i}\}$ & $\{\mathcal{C}_{i+1}\}$ &  $x_{i+1} \oplus x_{i+4t-1}$\\ 
			&&& $\{\mathcal{C}_{i+P-1}\}$ &  $x_{i+1} \oplus x_{i+4t}$\\ 
			&&&$\{\mathcal{C}_{i+2},...,\mathcal{C}_{i+P-2}\}$ &  $x_{i+4t-1} \oplus x_{i+4t}$\\  \hline
			
			$k=2$ and $q=2$& $w_{g+1},$ for $g \in [0,t-1]$ &$ \bigcup_{j=1}^{4} \{\mathcal{C}_{i+4g+j}\}$ &  $\mathcal{C} \backslash \bigcup_{j=1}^{4} \{\mathcal{C}_{i+4g+j}\}$ & $x_{i+4g} \oplus x_{i+4g+2}$ \\ \cline{2-5}
			& $w_{t+1}$ & $\{\mathcal{C}_{i},\mathcal{C}_{i-1}\}$ & $\{\mathcal{C}_{i+1}\}$ &  $x_{i+4t-1} \oplus x_{i+4t}$\\ 
			&&&$\{\mathcal{C}_{i+2},...,\mathcal{C}_{i+P-2}\}$ &  $x_{i+4t} \oplus x_{i+4t+1}$\\  \hline
			
			$k=2$ and $q=3$& $w_{g+1},$ for $g \in [0,t-1]$ &$ \bigcup_{j=1}^{4} \{\mathcal{C}_{i+4g+j}\}$ &  $\mathcal{C} \backslash \bigcup_{j=1}^{4} \{\mathcal{C}_{i+4g+j}\}$ & $x_{i+4g} \oplus x_{i+4g+2}$ \\ \cline{2-5}
			& $w_{t+1}$ & $\{\mathcal{C}_{i-1},\mathcal{C}_{i-2}\}$ & $\{\mathcal{C}_{i}\}$ &  $x_{i+4t-1} \oplus x_{i+4t}$\\ 
			&&&$\{\mathcal{C}_{i+1},...,\mathcal{C}_{i+P-3}\}$ &  $x_{i+4t} \oplus x_{i+4t+1}$\\ 
			& $w_{t+2}$ & $\mathcal{C}_{i}$ &  $\{\mathcal{C}_{i+1}\}$ &  $x_{i+1} \oplus x_{i+4t+1}$\\ 
			&&& $\{\mathcal{C}_{i+P-1}\}$ &  $x_{i+1} \oplus x_{i+4t+2}$\\ 
			&&&$\{\mathcal{C}_{i+2},...,\mathcal{C}_{i+P-2}\}$ &  $x_{i+4t+1} \oplus x_{i+4t+2}$\\  \hline

			$k=P-3$ and $P$ even &$w_1$ & $\{\mathcal{C}_{j}\}$, where $j \in \{2g+1\}, g \in [0,\frac{P-2}{2}]$ & $\mathcal{C} \backslash \mathcal{C}_{j}$ & $x_j \oplus x_{j+2}$\\ \cline{2-5}
			
			&$w_2$ & $\{\mathcal{C}_{j}\}$, where $j \in \{2g\}, g \in [0,\frac{P-2}{2}]$ & $\mathcal{C} \backslash \mathcal{C}_{j}$ & $x_j \oplus x_{j+2}$\\ \hline
			
			$k=P-3$ and $P$ odd &$w_1$ & $\{\mathcal{C}_{j}\}$, where $j \in \{2g+1\}, g \in [0,\frac{P-3}{2}]$ & $\{\mathcal{C}_{u}\}$, where $u \in \{2g\}, g \in [0,\frac{P-1}{2}]$ & $x_j \oplus x_{j+2}$\\ 
			&&&$\{\mathcal{C}_{u}\}$, where $u \in \{P-2\}$ & $x_{j+1} \oplus x_{j+2}$\\ 
			&&&$\{\mathcal{C}_{u}\}$, where $u \in \{P-1\}$ & $x_j \oplus x_{j+1}$\\ \cline{2-5}
			
			&$w_1$ & $\{\mathcal{C}_{j}\}$, where $j \in \{2g\} \cup \{P-2,P-1\}, g \in [0,\frac{P-1}{2}]$ & $\{\mathcal{C}_{j}\}$, where $j \in \{2g+1\}, g \in [0,\frac{P-3}{2}]$ & $x_j \oplus x_{j+2}$\\ \hline
						
		\end{tabular}
	\end{center}
	\caption{Table that illustrates the uniqueness of the messages decoded by the clients for $k=1, P-3$.}
	\label{tab11}
\end{table*}
Table \ref{tab10} illustrates the message decoded by each client for some values of $k$. It is illustrated in table \ref{tab11} that each client doesn't get more than one message for those values of $k$ (for the same reason as in Case 1).

\begin{exmp}
	Let us take an example where $k=2$ and $P$ even. Let $P=4$ and $k=2$. Let $i=1$. The coded symbol obtained is $w_1 = x_0 \oplus x_2 $. The clients in $\{\mathcal{C}_0,\mathcal{C}_3\}$ get $x_0$ and those in $\{\mathcal{C}_2, \mathcal{C}_3\}$ get $x_3$ from $w_1$.
\end{exmp}
\begin{exmp}
	Consider an example where $k=P-3$ and $P$ even. Let $P=8$ and $k=5$. The coded symbols obtained are $w_1 = x_0 \oplus x_2 \oplus x_4 \oplus x_6$ and $w_2= x_1 \oplus x_3 \oplus x_5 \oplus x_7$. The clients $\mathcal{C}_0$ gets $x_1$,  $\mathcal{C}_2$ gets $x_3$,  $\mathcal{C}_4$ gets $x_5$ and  $\mathcal{C}_6$ gets $x_7$ from $w_2$. The clients $\mathcal{C}_1$ gets $x_2$,  $\mathcal{C}_3$ gets $x_4$,  $\mathcal{C}_5$ gets $x_6$ and  $\mathcal{C}_7$ gets $x_8$ from $w_1$. None of the clients $\{\mathcal{C}_{j}\}$, where $j \in \{0,2,4,6\}$ can decode any message from $w_1$ as $x_j \oplus x_{j+2}$ is present in $w_1$, where $x_j $ and $x_{j+2}$ are not available as side information with the client  $\{\mathcal{C}_{j}\}$. Similarly none of the clients  $\{\mathcal{C}_{j}\}$, where $j \in \{1,3,5,7\}$ can decode any message from $w_2$ as $x_j \oplus x_{j+2}$ is present in $w_2$. Hence each client can decode only one message.
\end{exmp}

What is left is for 
\begin{enumerate}
	\item $k=1$ and $P$ odd,
	\item $k=P-2$ and $P$ odd.
\end{enumerate}

\subsection{Non existence of codes}
We will show that for both the cases - $k=1$ and $k=P-2$, where $P$ is odd, there doesn't exist any code where each client gets exactly one message. 

$k=1$ and odd $P$: If the coded symbol transmitted has more than two XORed messages, none of the clients can decode any message from that since the cardinality of side information is one. Hence coded symbols transmitted can have atmost two XORed messages.

Let us take the case where we initially transmit some message uncoded. If we transmit any message $x_i$ uncoded, all other clients except $\mathcal{C}_{i+1}$ get $x_i$ as they do not have $x_i$ as side information. Since the clients in $\mathcal{C}_{i+1}$ have only one side information $x_i$, to retrieve any message they do not have, we need to transmit either $x_j$ or $x_i \oplus x_j$, for some $x_j \in \mathcal{M} \backslash x_i$. All other clients except $\mathcal{C}_{j+1}$ get $x_j$ as they do not have $x_j$ as side information if we transmit $x_j$. Hence some clients get both $x_i$ and $x_j$. So this is ruled out. If we transmit $x_i \oplus x_j$, since $x_i$ is already transmitted, all the clients can decode $x_j$ from this. Hence this is also ruled out. In short, we cannot transmit any messages uncoded.

The only possibility left is to transmit XOR of two messages. If we transmit XOR of two messages, say $x_{i_1} \oplus x_{j_1}$, the clients in $\mathcal{C}_{i_1+1}$ can decode $x_{j_1}$ and the clients in $\mathcal{C}_{j_1+1}$ can decode $x_{i_1}$. None of the other clients can decode anything. Following that if we transmit some message $x_{i_2}$ uncoded, all the clients except $\mathcal{C}_{i_2+1}$ get $x_{i_2}$ (Clients in $\mathcal{C}_{i_2+1}$ already have this as side information). Hence either the clients in $\mathcal{C}_{i_1+1}$ or the clients in $\mathcal{C}_{j_1+1}$ can decode $x_{i_2}$. So we cannot transmit any messages uncoded and instead we transmit $x_{i_2} \oplus x_{j_2}$. If $x_{i_2}$ is same as $x_{i_1}$, the clients in $\mathcal{C}_{i_1+1}$ will get $x_{j_2}$ apart from $x_{j_1}$. Similarly $\mathcal{C}_{j_1+1}$ will get $x_{j_2}$ apart from $x_{i_1}$ if $x_{i_2}$ is same as $x_{j_1}$, $\mathcal{C}_{j_1+1}$ will get $x_{i_2}$ apart from $x_{i_1}$ if $x_{j_2}$ is same as $x_{j_1}$ and $\mathcal{C}_{i_1+1}$ will get $x_{i_2}$ apart from $x_{j_1}$ if $x_{j_2}$ is same as $x_{i_1}$. Hence $x_{i_1},x_{i_2}, x_{j_1}$ and $x_{j_2}$ are all different messages. Hence if we transmit $x_{i_2} \oplus x_{j_2}$, the clients in $\mathcal{C}_{i_2+1}$ can decode $x_{j_2}$ and the clients in $\mathcal{C}_{j_2+1}$ can decode $x_{i_2}$. We continue transmitting $x_{i_3} \oplus x_{j_3}, x_{i_4} \oplus x_{j_4},...,x_{i_{\frac{P-1}{2}}} \oplus x_{j_{\frac{P-1}{2}}}$, where each of the messages are distinct. With these coded transmissions $P-1$ set of clients in $\mathcal{C}$ are satisfied. There exists a set of clients $\mathcal{C}_{i_r}$ whose requirement is not met by those transmissions. The side information of those clients is $x_{i_r-1} = \mathcal{M} \backslash ((\bigcup_{l=1}^{\frac{P-1}{2}}x_{i_l}) \cup (\bigcup_{l=1}^{\frac{P-1}{2}}x_{j_l}))$. Hence we need to transmit either $x_j$ or $x_j \oplus x_{i_r-1}$, where $x_{j} \in ((\bigcup_{l=1}^{\frac{P-1}{2}}x_{i_l}) \cup (\bigcup_{l=1}^{\frac{P-1}{2}}x_{j_l}))$. If we transmit $x_j$ and if $x_j = x_{i_l},$ for some $l \in [1,\frac{P-1}{2}]$, then the clients in $\mathcal{C}_{i_l+1}$ get $x_{i_r+1}$ apart from $x_{i_l}$ and if $x_j = x_{j_l},$ for some $l \in [1,\frac{P-1}{2}]$, then the clients in $\mathcal{C}_{j_l+1}$ get $x_{i_r+1}$ apart from $x_{j_l}$. Hence we cannot transmit $x_j$. If we transmit $x_j \oplus x_{i_r-1}$ also, the above problem arises. Hence we cannot do that also.
				
To summarize, we cannot find any code such that all the clients get exactly one message.

$k=P-2$ and $P$ odd: We randomly pick some message $x_{i_1}$. It satisfies the requirements of the clients in $\mathcal{C}_{i_1-1}$ and $\mathcal{C}_{i_1}$. In general a message $x_{a}$ satisfies the requirements of exactly two set of clients - $\mathcal{C}_{a-1}$ and $\mathcal{C}_{a}$. Now we choose a message $x_{i_2}$ such that it is in the side information of $\mathcal{C}_{i_1-1}$ and $\mathcal{C}_{i_1}$. The message $x_{i_2}$ satisfies the requirements of the clients $\mathcal{C}_{i_2-1}$ and $\mathcal{C}_{i_2}$. We keep on choosing messages such that all the previously selected clients requirement is met even if we choose a pick a new message. We do this until no more messages can be picked. Assume $r$ messages are picked up, say $x_{i_1},x_{i_2},...,x_{i_r}$. If we XOR these $r$ messages and send, it satisfies the requirements of even number of clients - $\bigcup_{a=1}^{r} (\mathcal{C}_{i_a-1} \cup \mathcal{C}_{i_a})$. Since the cardinality of the set $\mathcal{C}$ is odd, there exists a set of clients $\mathcal{C}_{l} \notin \bigcup_{a=1}^{r} (\mathcal{C}_{i_a-1} \cup \mathcal{C}_{i_a})$. In order to satisfy the requirement of this set of clients, we need to transmit either $x_{j_1}$ or XOR of $x_{j_1}$ with some of the side information of the clients in $\mathcal{C}_{l}$, where $x_{j_1} \in \{x_{i_l},x_{i_l+1}\}$. If we transmit $x_{j_1}$ and if $x_{j_1}= x_{i_l}$, the clients in $\mathcal{C}_{i_l-1}$ get $x_{i_l}$ apart from $x_{i_l-1}$, since $x_{i_l-1} \in \{x_{i_1},x_{i_2},...,x_{i_r}\}$. If we transmit $x_{j_1}$ and if $x_{j_1}= x_{i_l+1}$, the clients in $\mathcal{C}_{i_l+1}$ get $x_{i_l+1}$ apart from $x_{i_l+2}$, since $x_{i_l+2} \in \{x_{i_1},x_{i_2},...,x_{i_r}\}$. Hence we cannot transmit $x_{j_1}$ alone. Now, if we transmit XOR of $x_{j_1}$ with some of the side information of the clients in $\mathcal{C}_{l}$ also the same argument holds. Hence we cannot do that also. Hence it is not possible to find a code such that all the clients get exactly one message for this case.

%Maximum

\subsection{Optimality}

In this section we provide the lower bound on the minimum number of scalar transmissions required for PICOD with consecutive side information. 

Let $P= t(P-k) +q$. 

For $q=0$, with one transmission each client gets one message they have demanded, for any value of $P$. The coded symbol is $$w_1 = \bigoplus_{g=0}^{t-1}x_{g(P-k)}$$. The client $\mathcal{C}_j$, where $j \in [(g)(P-k)+1,(g+1)(P-k)],g \in [0,t-1]$ gets  $x_{(g+1)(P-k)}$ from $w_1$.

For $q \neq 0$, we will prove in the coming part that with one transmission it is not possible for all the clients to retrieve the message they demanded. One transmission implies either sending a message uncoded or sending one coded symbol. If we send some message $x_i$ uncoded, then the client $\mathcal{C}_{i+1}$ doesn't get any message it doesn't have. Hence we cannot send any message uncoded. Considering the transmission of a coded symbol. Let us assume that it is possible for all the clients to retrieve the messages they demanded with one coded transmission. Let us start building up such a coded symbol. Start with a random message $x_i$. We cannot pick any of the messages in $\{x_{i+1},...,x_{i+P-k-1}\}$ to add to the coded symbol as the client $\mathcal{C}_{i}$ won't be able to decode any message if we add any one of them. We have to pick $x_{i+P-k}$, otherwise the client $\mathcal{C}_{i+1}$ won't be able to decode any message. Continuing the same argument we pick the messages $x_{i+g(P-k)}$, where $g \in [2,t-1]$. Next we have to pick $x_{i+t(P-k)}$, else the client $\mathcal{C}_{i+(t-1)(P-k)+1}$ won't be able to decode any message. But if we pick $x_{i+t(P-k)}$, the client $\mathcal{C}_{i+(t)(P-k)}$ won't be able to decode any message as both $x_i$ and $x_{i+t(P-k)}$ will be there in the coded symbol and both are not available as side information. Hence with one coded transmission it is impossible for all the clients to decode the message they demanded. So minimum number of coded transmissions required is two.

Hence for the second extreme case discussed in this section, the code provided is optimal.

\section{Total number of messages decoded by effective clients is maximized.}
\label{sec:case2}
In this section we will provide the code for the second extreme case where the effective clients get maximum number of messages.

Note: Any receiver $\mathcal{R}_j$ decodes atmost one message from any coded symbol $w_n$ transmitted. If it can decode two messages from $w_n$, say $x_a$ and $x_b$, then both the messages must be wanted by $\mathcal{R}_j$ and both have to be there in $w_n$. Since XOR of $x_a$ and $x_b$ is present in $w_n$, we cannot decode $x_a$ and $x_b$ individually. Hence $\mathcal{R}_j$ can decode atmost one message from the coded symbol $w_n$. 

So, from two coded symbols, a receiver can decode atmost two messages. Our objective is to find coded symbols, of optimal length, such that maximum receivers will get maximum messages from the coded symbols.

Let $P= (P-k)t_1 + q_1$ and $k+q_1 = (P-k)t_2 + q_2$. 

Let $b= \min{\{q_2, P-k-q_2\}}$.

The code construction is as follows. 
\begin{itemize}
	\item Pick some integer $i \in [0,P-1]$.
	\item For $P > 3k$,
	\begin{align*}
	w_1&=x_i.&\\
	w_2&=x_j, &\text{where $j$ can be any integer in the set}\\&& \text{$\{i+k+1,i+k+2,...,i+P-k+1\}$}.\\
	\end{align*}
	\item For $P \leq 3k$,
	\begin{itemize}
	\item If $q_1=0,$ one coded symbol $w_1$ is obtained, where
	$$w_1= \bigoplus_{j=0}^{t_1-1} x_{i+j(P-k)}.$$
	\item If $q_2=0$ or $b =q_2$ two coded symbols $w_1$ and $w_2$ are obtained, where
		\begin{align*}
			w_1&=\bigoplus_{j=0}^{t_1} x_{i+j(P-k)}.&\\
			w_2&=x_i \bigoplus_{j=0}^{t_2-1} x_{i+q_1+j(P-k)}.\\
		\end{align*}
		\item If $b=P-k-q_2$, two coded symbols $w_1$ and $w_2$ are obtained, where $w_1$ and $w_2$ are given below. 
			\begin{align*}
		w_1&=\bigoplus_{j=0}^{t_1} x_{i+j(P-k)}.&\\
		w_2&=x_i \oplus x_{i+k} \bigoplus_{j=0}^{t_2-1} x_{i+q_1+j(P-k)}.\\
		\end{align*}
	\end{itemize}
	\end{itemize}

\begin{table*}
	\begin{center}
		\begin{tabular}{ | p{2.1cm} | p{6.8cm} | p{4.3cm}  |}
			\hline
			
			\textit{}& \textit{The set of all receivers who get minimum messages (exactly one message) from the coded symbols}  &  \textit{Total number of effective clients who get exactly one message} \\ \hline
				$P > 3k$ & $\{\mathcal{C}_{i+1},\mathcal{C}_{i+2},...,\mathcal{C}_{i+k} \cup \mathcal{C}_{j+1},\mathcal{C}_{j+2},...,\mathcal{C}_{j+k}\}$ & $2k$\\ \hline
			
			$P \leq 3k$ and $b=q_2$ & $\{\mathcal{C}_{i+k-b+1},\mathcal{C}_{i+k-b+2},...,\mathcal{C}_{i+k}\}$ & $q_2$\\ \hline
			
			$P \leq 3k$ and $b=P-k-q_2$ & $\{\mathcal{C}_{i+q_1+(t_2-1)(P-k)-b+1},\mathcal{C}_{i+q_1+(t_2-1)(P-k)-b+2},...,$ & $P-k-q_2$\\ 
			&$\mathcal{C}_{i+q_1+(t_2-1)(P-k)}\}$&\\
			\hline

		\end{tabular}
	\end{center}
	\caption{Table that illustrates the receivers who get least number of messages.}
	\label{tab7}
\end{table*}

Table \ref{tab7} illustrates the effective clients who get minimum number of messages with the coded transmissions. If $q_1 =0$, then each client gets exactly one message from one coded transmission. If $q_2=0$, every client gets exactly two messages from the two coded transmission. For all other cases, from the two coded transmissions, $P-b$ number of effective clients get two messages. All other clients can decode only one message.

\begin{exmp}
	Let $P=10,k=6$. Here, $t_1=2,q_1=2,t_2=2$ and $q_2=0$. $b=q_2=0$
	 Let $i=0$. The coded symbols obtained are $w_1 = x_0 \oplus x_4 \oplus x_6$ and $w_2 = x_0 \oplus x_2 \oplus x_6 $. All the clients get two messages from the two coded transmissions.  
\end{exmp}

\section{Constrained Pliable Index Coding }
\label{Sec:Case3}
 In this section, we provide index code for a class of PICOD where each client has any $k$ consecutive messages as side information, under a c-constraint, i.e., each message is decoded by atmost c clients demanding that message. We consider the case where $c\geq k$.

For $c \geq k$, if $P-k \leq c$, then use the index code for the second extreme case given in the previous section.

For $c \geq k$, if $P-k > c$, the code construction is as follows.. Let $P-2k-1 = kt +q$.

\begin{itemize}
	\item Pick some integer $i \in [0,P-1]$.
	\item For even $t$, $\lfloor \frac{t}{2} \rfloor +2$ coded symbols are obtained, where
    
    \begin{itemize}
    	\item for each $j = [0,\lfloor \frac{t}{2} \rfloor]$, a coded symbol\\ $w_{j+1} = x_{i+2jk} \oplus x_{i+(2j+1)k}$ is obtained.
    	\item for $j=\lfloor \frac{t}{2} \rfloor +1$, a coded symbol\\ $w_{j+1} = x_{i+2jk} \oplus x_{i+(2j-1)k-1}$ is obtained.
    \end{itemize}
	\item For odd $t$, $\lfloor \frac{t}{2} \rfloor +2$ coded symbols are obtained, where

\begin{itemize}
	\item for each $j = [0,\lfloor \frac{t}{2} \rfloor+1]$, a coded symbol\\ $w_{j+1} = x_{i+2jk} \oplus x_{i+(2j+1)k}$ is obtained.

\end{itemize}

\end{itemize}

  \begin{exmp}
  	Let $P=9,k=4$ and $c=4$. $P-2k-1=0$. Hence $t=0$. Let $i=0$. The coded symbols obtained are $w_1 =x_0 \oplus x_4$ and $w_2=x_3 \oplus x_8$. With this transmissions, the message $x_3$ is decoded by the clients $\{\mathcal{C}_0,\mathcal{C}_1,\mathcal{C}_2,\mathcal{C}_3\}$, the message $x_4$ by $\{\mathcal{C}_4,\mathcal{C}_1,\mathcal{C}_2,\mathcal{C}_3\}$, the message $x_8$ by $\{\mathcal{C}_4,\mathcal{C}_5,\mathcal{C}_6,\mathcal{C}_7\}$ and the message $x_0$ by $\{\mathcal{C}_5,\mathcal{C}_6,\mathcal{C}_7,\mathcal{C}_8\}$. Hence all the messages are decoded by four clients which is equal to $c$.
  \end{exmp}

\section{Conclusion}
\begin{itemize}
	\item We provide index code for two extreme classes of PICOD - for the class where each client gets exactly one desired message and  for a class where total number of messages decoded by the effective clients is maximized.
	\item We also provide index code for c-constrained pliable index coding with consecutive side information. 
\end{itemize}

\section*{Acknowledgment}
This work was supported partly by the Science and Engineering Research Board (SERB) of Department of Science and Technology (DST), Government of India, through J. C. Bose National Fellowship to B. Sundar Rajan.

\end{document}